\documentclass[conference]{IEEEtran}
\IEEEoverridecommandlockouts
\usepackage{cite}
\usepackage{amsmath,amssymb,amsfonts}
\usepackage{algorithmic}
\usepackage{graphicx}
\usepackage{textcomp}
\usepackage{xcolor}
\def\BibTeX{{\rm B\kern-.05em{\sc i\kern-.025em b}\kern-.08em
    T\kern-.1667em\lower.7ex\hbox{E}\kern-.125emX}}

\usepackage[ruled,vlined,linesnumbered]{algorithm2e}
\usepackage{setspace}
\usepackage{xspace}
\usepackage{enumitem}
\usepackage{url}
\usepackage{multirow}
\usepackage{hyperref}
\hypersetup{hidelinks} 
\usepackage{booktabs}
\usepackage{tcolorbox}
\usepackage{bbm}
\usepackage{balance}

\def \projectName {\textsc{SeaM}\xspace}

\begin{document}

\title{Reusing Deep Neural Network Models through Model Re-engineering}

\author{\IEEEauthorblockN{Binhang Qi*\thanks{* Also with Beijing Advanced Innovation Center for Big Data and Brain Computing, Beihang University, Beijing 100191, China.}
}
\IEEEauthorblockA{\textit{SKLSDE Lab, Beihang University} \\
Beijing, China \\
binhangqi@buaa.edu.cn}

\\

\IEEEauthorblockN{Hongyu Zhang}
\IEEEauthorblockA{\textit{Chongqing University} \\
 Chongqing, China \\
hyzhang@cqu.edu.cn}

\and

\IEEEauthorblockN{Hailong Sun*\textsuperscript{\textdagger}\thanks{\textsuperscript{\textdagger} Corresponding authors: Hailong Sun and Xiang Gao.}}
\IEEEauthorblockA{\textit{SKLSDE Lab, Beihang University} \\
Beijing, China \\
sunhl@buaa.edu.cn}

\\

\IEEEauthorblockN{Zhaotian Li*}
\IEEEauthorblockA{\textit{SKLSDE Lab, Beihang University} \\
Beijing, China \\
lizhaotian@buaa.edu.cn}

\and

\IEEEauthorblockN{Xiang Gao*\textsuperscript{\textdagger}}
\IEEEauthorblockA{\textit{SKLSDE Lab, Beihang University} \\
Beijing, China \\
xiang\_gao@buaa.edu.cn}

\\

\IEEEauthorblockN{Xudong Liu*}
\IEEEauthorblockA{\textit{SKLSDE Lab, Beihang University} \\
Beijing, China \\
liuxd@act.buaa.edu.cn}
}

\maketitle
\pagestyle{plain}
\thispagestyle{plain}

\begin{abstract}
Training deep neural network (DNN) models, which has become an important task in today's software development, is often costly in terms of computational resources and time.
With the inspiration of software reuse, building DNN models through reusing existing ones has gained increasing attention recently.
Prior approaches to DNN model reuse have two main limitations: 1) reusing the entire model, while only a small part of the model's functionalities (labels) are required, would cause much overhead (e.g., computational and time costs for inference), and 2) model reuse would inherit the defects and weaknesses of the reused model, and hence put the new system under threats of security attack.
To solve the above problem, we propose \projectName, a tool that re-engineers a trained DNN model to improve its reusability.
Specifically, given a target problem and a trained model, \projectName utilizes a gradient-based search method to search for the model's weights that are relevant to the target problem.
The re-engineered model that only retains the relevant weights is then reused to solve the target problem.
Evaluation results on widely-used models show that the re-engineered models produced by \projectName only contain 10.11\% weights of the original models, resulting 42.41\% reduction in terms of inference time.
For the target problem, the re-engineered models even outperform the original models in classification accuracy by 5.85\%.
Moreover, reusing the re-engineered models inherits an average of 57\% fewer defects than reusing the entire model.
We believe our approach to reducing reuse overhead and defect inheritance is one important step forward for practical model reuse.
\end{abstract}


\section{Introduction}
\label{sec:intro}
Software reuse is the process of using existing software artifacts that would be otherwise created from scratch~\cite{software_reuse_1,software_reuse_2,software_reuse_3}, which is widely deemed essential to improve software quality and development productivity. 
Instances of software reuse include the reuse of software libraries, components, APIs, etc. 
As today's software systems are increasingly incorporating AI techniques (e.g., deep learning),  training DNN models has become an important task in the software development lifecycle. 
However, training DNN models is often known to be 
very costly, especially for models with billions of parameters and large datasets.
To solve this problem, with the inspiration of software reuse, the software engineering community is paying more attention to DNN model reuse~\cite{icse21discriminiate, ji2018model, modeldiff, ReMos, nnmodularity2022icse,fse2020modularity, qi2022patching}.

A trained model can be directly reused if it fits %
the {target problem} domain. %
However, reusing entire trained models may cause large %
overhead (e.g., inference time).
Just like traditional software libraries which implement a large number of functions, a trained model may also have multiple functionalities (e.g., classification for multiple categories).
When reusing a trained model, often only part of functionalities are required to solve the target problem.
For instance, Google Vision API provides the service of  multi-class classification with around 20,000 classes, but not all classes are necessary in practical scenarios. Suppose that a developer needs to build a fire alarm application~\cite{fireapp} for determining whether a given image indicates ``fire''.
Although only two classes (``fire'' and ``non-fire'') are needed, if the developer directly invokes Google Vision API, all the 20,000 classes will be involved, which can incur much inference overhead caused by the unnecessary weights/neurons in the underlying DNN model.

A model trained to solve a similar problem can also be indirectly reused via transfer learning~\cite{transfer_nips,devlin2018bert}.
Transfer learning consists of taking relevant features learned on a similar problem and optionally fine-tuning the trained model using the dataset of the target problem. %
Although effective in classification accuracy and training efficiency, reusing trained models may inherit their defects %
~\cite{dlfuzz,deepxplore,defect3}.
It has been shown that AI models are notoriously brittle to small perturbations on input data~\cite{huang2011adversarial, deepxplore},
which allows attackers to craft adversarial examples for malicious attacks.
When reusing a model, the weakness of a trained model can be inherited, and hence putting the system under the threats of adversarial attacks.

To address the weaknesses of existing model reuse methods, one idea is to only reuse some parts of a trained model (e.g., by eliminating some weights or neurons) that are relevant to the target problem, as the weaknesses correlate with the weights of a trained model~\cite{rezaei2019target, ReMos}.
Identifying the relevant weights/neurons can be achieved with the fundamental concept of \textit{re-engineering} in software engineering
~\cite{rosenberg1996software,chikofsky1990reverse}, which aims to improve software maintainability and reusability by
enhancing or altering existing software.
Borrowing the idea of software re-engineering, we propose \emph{model re-engineering} for DNN models, which searches for the target problem-related weights with the guidance of target problem-related metrics (e.g., classification accuracy) and removes irrelevant weights from an \textit{original model} (i.e., trained model), resulting in a re-engineered model.
When solving a certain problem through direct or indirect reuse, the re-engineered model, which retains only relevant weights to certain functionalities (e.g., a part of classes in classification), is reused, hence reducing the reuse overhead and mitigating the defect inheritance. %

Existing work, including model modularization~\cite{nnmodularity2022icse,fse2020modularity,qi2022patching} and model slicing~\cite{ReMos}, has preliminarily explored the idea of reusing part of trained models based on neuron activation and %
neuron coverage~\cite{deepxplore, dlfuzz}.
For instance, relying on neuron coverage, model slicing~\cite{ReMos} first computes the relevance between weights and the target problem, then deletes the irrelevant weights.
Unfortunately, due to the lack of interpretability of DNN models, the effectiveness of using neuron coverage is still questionable~\cite{neuron_activation,li2019structural}.
The neuron coverage-based work~\cite{nnmodularity2022icse,fse2020modularity,ReMos} is not accurate enough in identifying relevant weights and hence prefers to be conservative in removing weights, i.e., only a small number of weights are removed to avoid removing relevant weights.
Therefore, the models %
obtained with the existing approaches~\cite{nnmodularity2022icse,fse2020modularity,ReMos} will retain lots of irrelevant weights or neurons, having the limitations of reuse overhead and defect inheritance.
CNNSplitter~\cite{qi2022patching} introduces the first search-based approach for modularizing CNNs. As CNNSplitter achieves modularization by searching for relevant convolution kernels with a genetic algorithm, this approach cannot be applied directly to other neural networks, such as the fully connected neural networks.

In this paper, we propose \projectName, a \textbf{Sea}rch-based \textbf{M}odel re-engineering approach that can accurately identify relevant weights and hence removes as many irrelevant weights as possible.
Different from the neuron coverage-based approaches~\cite{nnmodularity2022icse,fse2020modularity,ReMos}, \projectName is directly guided by the target problem-related metrics, e.g., classification accuracy, to search for the relevant weights.
Moreover, \projectName applies a gradient-based search method to identify relevant weights, which is more general and efficient than CNNSplitter~\cite{qi2022patching}.
Specifically, \projectName consists of three components: search space, performance estimation strategy, and search strategy.
The search space consists of the masks of all candidate re-engineered models. The mask of a candidate records which weights of the original model should be retained or removed. %
The performance estimation strategy defines the objective function of the search as the weighted sum of the candidate's weight retention rate and its cross-entropy loss on the target problem's dataset (denoted as target dataset).
The objective function is used to evaluate a candidate's performance, and the objective function value is sent to the search strategy to guide the next round of search.
The search strategy applies a gradient-based search method to explore the search space efficiently. 
In each search round, the search strategy finds a candidate with better performance by minimizing the objective function value. %
\projectName performs the search and estimation processes iteratively, and stops %
when the objective function value converges.
The candidate with the minimal objective function value will be regarded %
as the resultant re-engineered model.
The re-engineered model can be reused directly, or indirectly via fine-tuning, which helps reduce reuse overhead and lower the risk of defect inheritance while achieving comparable performance (e.g., classification accuracy) to the original model.

We evaluate \projectName using four representative CNN models on eight widely-used datasets. 
The experimental results first demonstrate that \projectName can accurately identify relevant weights and thus remove a large number of irrelevant weights. 
On average, a re-engineered model contains 89.89\% fewer weights than the original model, and outperforms the original model by 5.85\% in classification accuracy.
Moreover, reusing a re-engineered model incurs less reuse overhead than reusing an original model, e.g., the average reduction in time cost for inference is 42.41\%. Regarding defect inheritance, reusing the re-engineered model inherits an average of 57\% fewer defects than reusing the original model.

The main contributions of this work are as follows:  
\begin{itemize}[leftmargin=*]
    \item We propose the notion of \textit{model re-engineering}, which re-engineers a trained deep learning model %
    to improve its reusability.
    \item We propose a search-based model re-engineering approach named \projectName, which can accurately identify the weights relevant to a target problem and hence allows the re-engineered model to retain as few irrelevant weights as possible. \projectName can reduce the reuse overhead and lower the risk of defect inheritance in model reuse.
    \item We conduct extensive experiments using four representative CNN models on eight widely-used datasets. The results show that \projectName can remove a large number of irrelevant weights from the original models. %
    Also, the experiments demonstrate the effectiveness of \projectName in overcoming the limitations of existing approaches.%
\end{itemize}

\begin{figure}[t]
    \centering
    \includegraphics[width=\columnwidth]{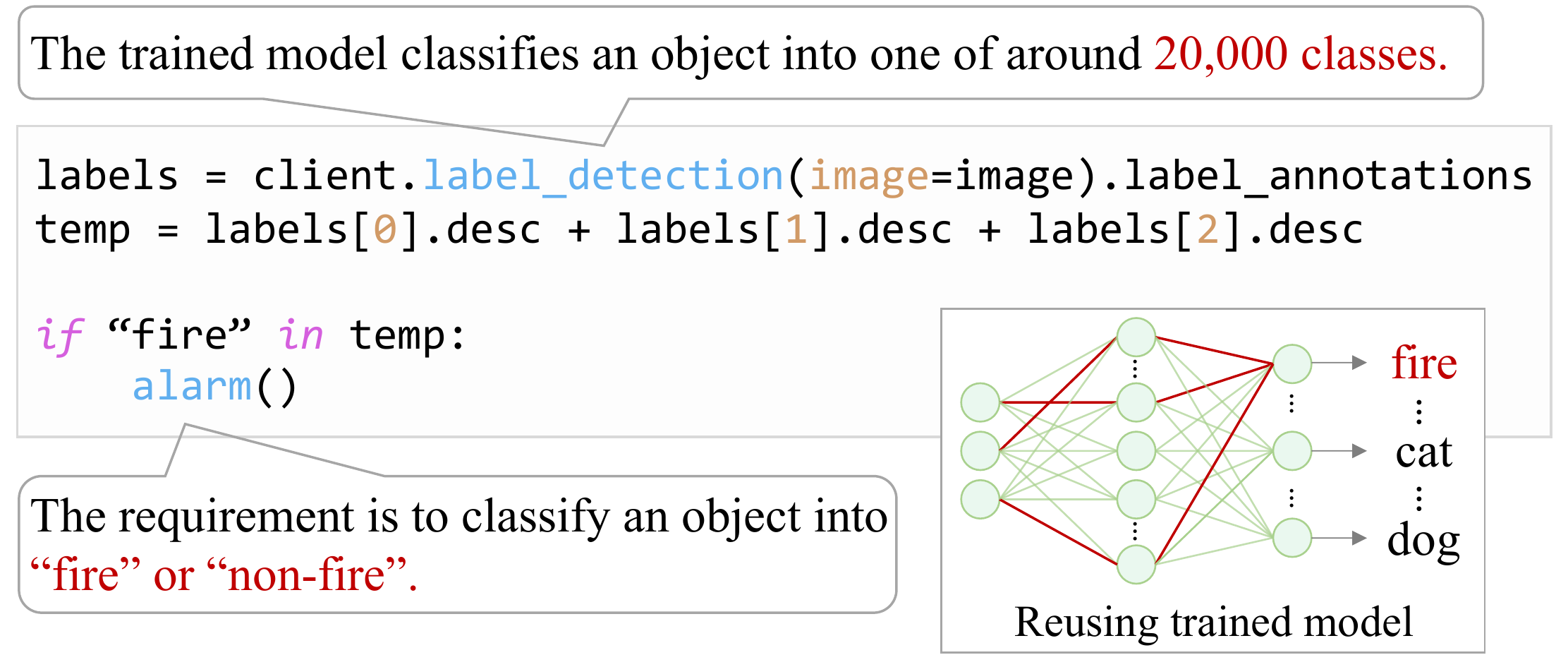}
    \caption{An example of direct model reuse.}%
    \label{fig:motivation_1}
\end{figure}

\section{Motivating Examples}
\label{sec:motivation}

Reusing a re-engineered model containing fewer irrelevant weights rather than an original model has several benefits. In this section, we introduce the applications and benefits of model re-engineering with two examples.

\subsection{Reducing reuse overhead in direct reuse}
\label{subsec:motivation_1}

When a trained model satisfies the requirement of a target problem, a common way of reuse is to reuse the entire trained model on the target problem directly.
However, there may be redundancy in the functionalities provided by the trained model~\cite{nnmodularity2022icse,mltest}.
Redundancy in a trained model's functionality implies redundant weights, which may incur significant \textit{reuse overhead}, including computational and time costs for inference, that is unnecessary for the target problem.

As shown in Figure \ref{fig:motivation_1}, a simple fire alarm application~\cite{fireapp} is used to illustrate the problem.
In this example, the developer reuses a trained model (by calling the Google \texttt{label\_detection} API) to classify an input image.
An alarm will be triggered if the top-3 classification labels returned by the trained model include the keyword ``fire''.
The requirement of the target problem is to classify an image into ``fire'' or ``non-fire'', while the reused trained model classifies an image into one of around 20,000 classes.
As different weights could recognize features of different classes~\cite{yamashita2018convolutional,bau2020understanding}, only a few relevant weights recognize the features of ``fire''.
However, when reusing the trained model for inference, a lot of irrelevant weights are loaded into memory and involved in computation to produce intermediate results, incurring memory, computational, and time costs.

The example demonstrates that the requirement of a target problem may be only a small part of a trained model's functionality.
Model re-engineering can remove part of the original model's weights irrelevant to the target problem and allows developers to reuse only the relevant weights.
In this example, 
the weights irrelevant to the target problem are removed, resulting in a re-engineered model that only classifies ``fire'' and ``non-fire''.
Compared to directly reusing the trained model, reusing the re-engineered model containing fewer weights could reduce the reuse overhead.

\subsection{Mitigating defect inheritance in transfer learning}

\begin{figure}[t]
    \centering
    \includegraphics[width=\columnwidth]{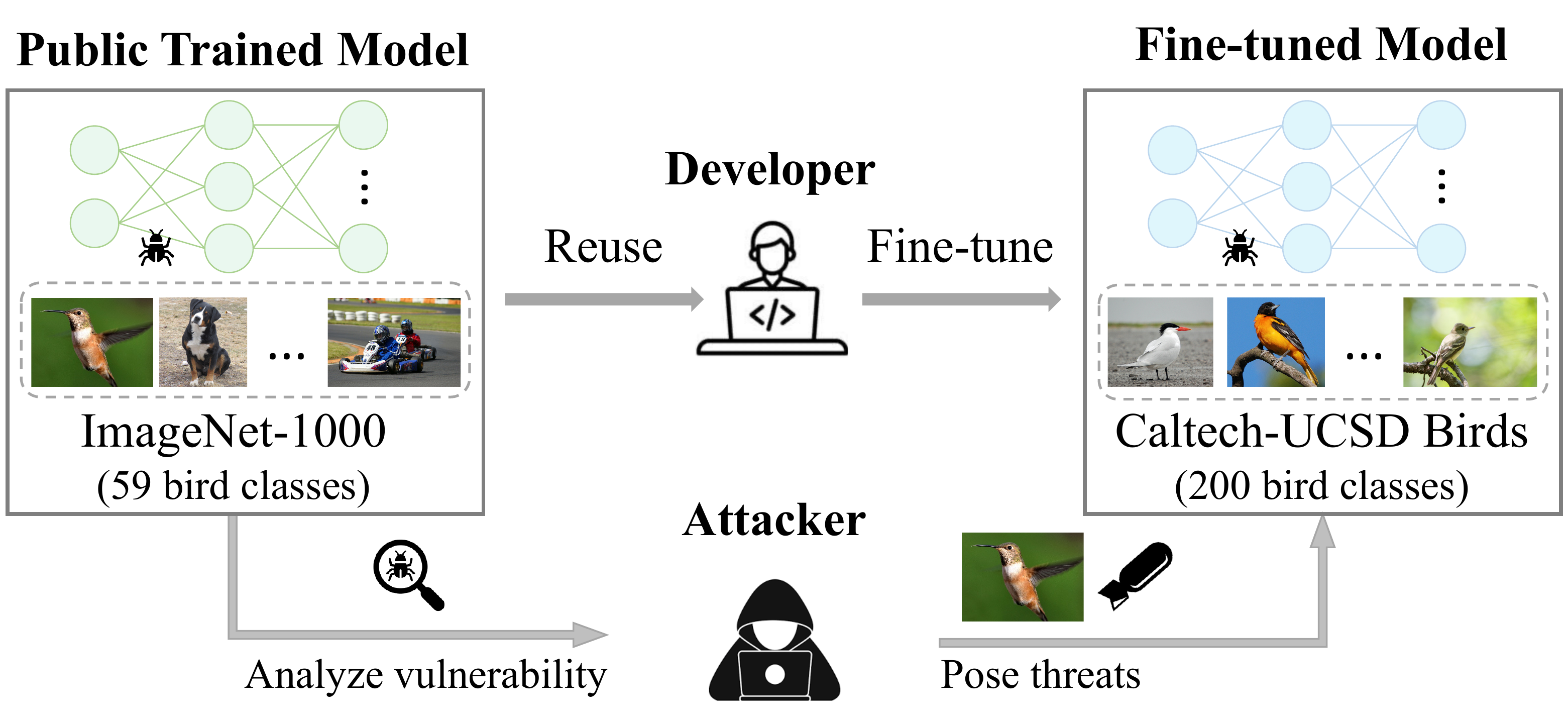}
    \caption{Fine-tuning a publicly available trained model. Inherited defects could be exploited by 
    attackers. }
    \label{fig:motivation_2}
\end{figure}

When a trained model cannot satisfy the requirement of a target problem, a common form of reuse is to transfer learning~\cite{transfer1,transfer2,transfer3}.
That is, a developer reuses a trained model and fine-tunes the trained model on the target dataset to build a fine-tuned model that satisfies the requirement.
This form of reuse is widely-used and effective; however, it faces the problem of defect inheritance~\cite{ReMos,defect1,defect2,defect3}.
An example shown in Figure \ref{fig:motivation_2} is used to illustrate the defect inheritance and potential attacks to face.
In this example, a public model trained on ImageNet~\cite{imagenet} can perform classification with 1000 classes (including 59 bird classes~\cite{van2015building}).
To build a model for classifying birds with 200 classes, a developer reuses the trained model and fine-tunes the trained model on the target dataset Caltech-UCSD Birds~\cite{tfdata_birds}.
During fine-tuning, most of the weights in the pre-trained model are retained in the fine-tuned model.
The adversarial examples that can fool the public trained model are still likely to be able to fool the fine-tuned model, which is called defect inheritance~\cite{ReMos,defect1,defect2,defect3}.

The major reason for defect inheritance is indiscriminate reuse~\cite{rezaei2019target, ReMos}.
Specifically, in conventional transfer learning, all the trained model's weights are reused, including both the relevant and the irrelevant ones to the target problem.
As the target dataset is usually not very large, fine-tuning will not have much effect on changing the weights irrelevant to the target problem. As a result, 
the defects %
are mostly inherited in the fine-tuned model~\cite{liu2019wealthadapt, ReMos}.%

Model re-engineering alters the original model by removing irrelevant weights, thus avoiding the inheritance of defects associated with these weights when the re-engineered model is reused.
In this example, a re-engineered model retains only the weights relevant to the features of ``bird''.
As a result, compared to reusing the original model, reusing the re-engineered model can reduce the defect inheritance while achieving comparable accuracy.

\section{Our Approach}
\label{sec:approach}
In this section, we introduce \projectName, a search-based approach to model re-engineering, which uses a gradient-based search method to find the target problem-related weights.

\begin{figure*}[t]
    \centering
    \includegraphics[width=17.5cm]{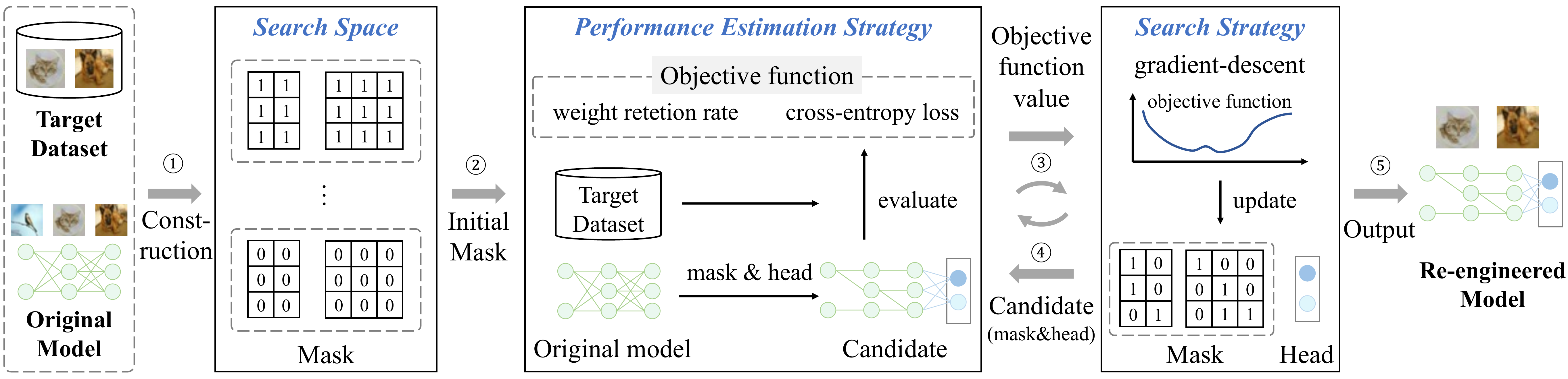}
    \caption{The workflow of model re-engineering with \projectName.}
    \label{fig:workflow}
\end{figure*}

\subsection{Overview}
As illustrated in Figure \ref{fig:workflow}, the workflow of \projectName consists of three components: \textit{search space}, \textit{performance estimation strategy}, and \textit{search strategy}. 
Given an original model (a 3-class classification in Figure \ref{fig:workflow}), which consists of three neural network layers with fifteen weights, and a target dataset (binary classification in Figure \ref{fig:workflow}), the model re-engineering process is summarized as follows:

(1) \textit{Construction of Search Space}: A re-engineered model selectively removes part of the original model's weights according to a \textit{mask}. 
A \textit{mask} is a bit vector $[0,1]^L$, where $L$ is the number of weights in the original model, and each bit represents whether the corresponding weight is removed. 
In total, there are $2^L$ candidate masks, each of which corresponds to a candidate re-engineered model.
Consequently, the search space consists of $2^L$ candidates.
The mask is initialized with all element values as 1, representing that all weights are retained initially.
The first and second steps along with the component Search Space in Figure \ref{fig:workflow} display the above process, where $L=15$ and the search space size is $2^{15}$.

(2) \textit{Performance Estimation}: Given a candidate mask, the performance estimation strategy first constructs a candidate re-engineered model by removing weights according to the mask and appending a \textit{head} as the output layer.
The \textit{head}, which is a fully connected layer, is used to enable the candidate to adapt to the target problem, i.e., adapt the original \textit{N}-classification model to the target \textit{K}-classification problem.
Then, the objective function is defined as the weighted sum of the \textit{weight retention rate} of the candidate and the \textit{cross-entropy loss} between the candidate's predictions and corresponding actual labels on the \textit{target dataset}.
The objective function is used to evaluate the performance of a candidate. 
The resulting objective function value will be fed back to the searching process to guide the next search round. %
The third step along with the component Performance Estimation Strategy in Figure \ref{fig:workflow} display the estimation process.

(3) \textit{Searching Candidates}: The search strategy applies a gradient-based search method to explore the search space with the guidance of the objective function. The gradient-based search method not only efficiently explores the huge search space, but also optimizes the head at the same time.
In each search round, the search strategy sends the updated mask and head as a new candidate to the performance estimation strategy.
The fourth step along with the component Search Strategy in Figure \ref{fig:workflow} display the search process, where the head has two neurons as the target problem is binary classification.

\projectName iterates the search and estimation processes.
When the objective function value converges, \projectName %
outputs the re-engineered model. %
In the example shown in Figure \ref{fig:workflow}, the re-engineered model retains 7 out of 15 weights of the original model and performs binary classification.
We present the technical details of each step in the following.

\subsection{Construction of Search Space}
\begin{figure}
    \centering
    \includegraphics[width=\columnwidth]{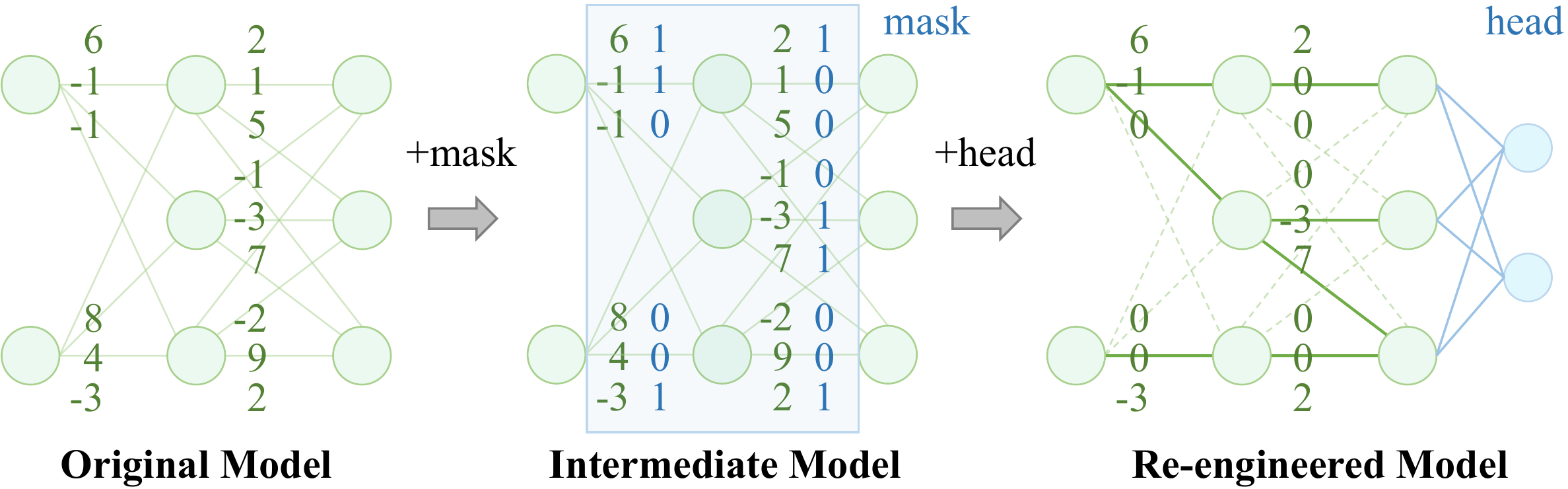}
    \caption{The construction of a re-engineered model using the mask and head. }
    \label{fig:removal}
\end{figure}

The goal of model re-engineering is to obtain a new model which retains only the target problem-related weights of the original model.
Model re-engineering is formulated as a problem of searching for a new model from all candidate models, which selectively removes part of the original model's weights. 
If the searched model retains only the target problem-related weights, it is regarded as the re-engineered model.
In this problem, the search space consists of all possible candidate re-engineered models.
To facilitate a technical solution to this problem in practice, a \textit{mask} that records which weights are removed and retained in a candidate is used to represent a candidate, thereby omitting unnecessary details of a candidate, such as Max-pooling and Dropout layers.
Consequently, in \projectName, the search space consists of all candidate masks.

Specifically, a mask is a bit vector $[0,1]^L$, where $L$ is the number of weights in the original model, and 0 (or 1) represents the corresponding weight removed (or retained). 
Figure \ref{fig:removal} illustrates the use of a mask to remove weights from the original model.
By multiplying the weights of the trained model with the mask, \projectName sets the values of irrelevant weights to zero and keeps the values of relevant weights. 
The weights with values set to zero are involved in the computation but have no effect on the prediction, thus achieving the effect of removing irrelevant weights.
Note that, after model re-engineering, the computation of a re-engineered model involving the weights with zero values could be eliminated by special libraries (e.g., DeepSparse~\cite{deepsparse}), which will be discussed in Section \ref{subsec:exp_result}.

After the construction of search space, a mask initialized to all element values of 1 is fed to the performance estimation strategy. That is, the starting point of the search is a candidate that retains all the original model's weights.

\subsection{Performance Estimation}
The search aims to find the optimal mask, which corresponds to a candidate re-engineered model that retains only the target problem-related weights and can classify well on the target problem.
To achieve the goal, the performance estimation strategy defines the objective function of the search as the weighted sum of \textit{weight retention rate} and \textit{cross-entropy loss}.
The weight retention rate can measure the number of weights retained by the candidate.
The cross-entropy loss on the target dataset can measure the classification performance of the candidate on the target problem.

Specifically, when evaluating a candidate's performance, \projectName first constructs the candidate, as the computation of cross-entropy loss requires running the candidate on the target dataset.
Figure \ref{fig:removal} illustrates the construction of a candidate re-engineered model.
\projectName first multiplies the weights of the original model with the mask to remove part of the original model's weights, resulting in an intermediate model.
As the output layer has three neurons, the intermediate model is still a model for 3-class classification. 
To adapt the candidate to the number of classes of the target problem, the head, a fully connected layer, is appended after the intermediate model as the output layer of the candidate.
The head is randomly initialized in the first search round and will be updated along with the mask in the subsequent rounds.
In this example, the head has two neurons, which transforms the 3-class prediction of the intermediate model to the binary prediction, allowing the candidate to adapt to the target problem.

After constructing the candidate, the cross-entropy loss $\mathcal{L}_{ce}$ between the candidate's predictions on the target dataset and the actual labels is computed as follows:
\begin{gather}
    \mathcal{L}_{ce}=-\sum_{i=1}^K t_i \log(P_i(\mathcal{M}, \mathcal{H})),
\end{gather}
where $K$ is the number of classes in the target problem, $\mathcal{M}$ and $\mathcal{H}$ are the mask and head, $P_i(\mathcal{M}, \mathcal{H})$ is the prediction for the $i$-th class by a candidate constructed with $\mathcal{M}$ and $\mathcal{H}$, and $t_i$ is the probability of the $i$-th class in the one-hot representation of the actual label, with a value of 0 or 1.
A lower cross-entropy loss indicates that the candidate retains more target problem-related weights and hence achieves higher classification accuracy on the target dataset.

The weight retention rate $\mathcal{L}_{wr}$ is computed directly from the mask:
\begin{gather}
    \mathcal{L}_{wr}=\frac{1}{L}\sum_{i=1}^L \mathcal{M}[i],
\end{gather}
where $L$ is the number of weights in the original model.
A lower weight retention rate indicates that the candidate retains fewer weights.
Based on $\mathcal{L}_{ce}$ and $\mathcal{L}_{wr}$, the objective function $\mathcal{O}$ is defined as follows:
\begin{gather}
    \mathcal{O}=\mathcal{L}_{ce} + \alpha \times \mathcal{L}_{wr}, \label{eq:loss}
\end{gather}
where $\alpha$ is a weighting factor and is empirically set to 1.0. %
To minimize $\mathcal{O}$, \projectName tends to search for a candidate that retains only the target problem-related weights, as this candidate can achieve the highest classification accuracy while retaining as few weights as possible.

\subsection{Searching Candidates}
Large models can have billions of parameters, resulting in super huge search space.
To explore the huge search space efficiently, our search strategy applies a gradient-based search method.
In each search round, the search strategy finds a new candidate with a smaller objective function value by gradient descent based on the objective function value of the candidate in the previous round.
That is, the mask is updated by descending the gradient as follows:
\begin{gather}
    \mathcal{M^{'}} = \mathcal{M} - \xi \times \nabla_{\mathcal{M},\mathcal{H}}\mathcal{O}, \label{eq:update} \\
    \nabla_{\mathcal{M},\mathcal{H}}\mathcal{O}=\nabla_{\mathcal{M}, \mathcal{H}}\mathcal{L}_{ce} + \alpha \times \nabla_{\mathcal{M}}\mathcal{L}_{wr},
\end{gather}
where $\xi$ is the learning rate, and $\mathcal{M^{'}}$ is the updated mask corresponding to a new candidate with a smaller objective function value.

When applying gradient descent to update a mask, it is important to note that gradient descent requires the search space to be continuous and differentiable~\cite{darts}, while the search space composed of masks is discrete and non-differentiable.
Inspired by DARTS~\cite{darts}, the search strategy attaches a continuous number to each element of the mask, which can be considered as the relevance of the weight to the target problem.
Then an indicator function $\mathbbm{1}_{(0, +\infty)}{:}X {\to} \{0,1\}$ is used to set the element values corresponding to the weights with relevance greater than zero to 1 and the other element values to 0.
As the relevance is continuous, the search strategy uses gradient descent to update the relevance and thus can update the mask.

After satisfying the condition that the search space is continuous, another problem is that the indicator function is non-differentiable at $x{=}0$, and the derivative of the indicator function equals 0 everywhere except at $x{=}0$.
This problem prevents the common backward propagation based on gradient descent from directly applying to update relevance~\cite{binaryNN,binarizedNN_2}.
To address the problem, the technique named straight-through estimator~\cite{ste,binarizedNN_2} is used to estimate the gradient of the indicator function, which directly uses the gradient of the previous neural network layer as the gradient of the current neural network layer.

The head is updated along with the mask by descending the gradient $\nabla_{\mathcal{M},\mathcal{H}}\mathcal{L}_{ce}$.
After updating the mask and head, the search strategy sends them as a new candidate to the performance estimation strategy and starts the next round of search after getting the objective function value.

\section{Experiments}
\label{sec:exp}
To evaluate the effectiveness of \projectName, in this section, we introduce the benchmarks and experimental setup as well as the experimental results.
Specifically, we evaluate \projectName by answering the following research questions:
\begin{itemize}[leftmargin=*]
    \item RQ1: How effective is our model re-engineering approach in reusing trained models?
    \item RQ2: Does reusing a re-engineered model incur less overhead than reusing the original model?
    \item RQ3: Does reusing the re-engineered model mitigate the defect inheritance? %

\end{itemize}

\subsection{Experimental Setup}
\label{subsec:setup}
\noindent \textbf{RQ1: How effective is our model re-engineering approach in reusing trained models?} Three representative CNN models are used in this research question, including VGG16~\cite{vgg}, ResNet20, and ResNet50~\cite{resnet}. 
The three CNN models are trained on three public classification datasets, including CIFAR-10~\cite{cifar10}, CIFAR-100~\cite{cifar10}, and ImageNet~\cite{imagenet}.
In total, there are five trained CNN models in this experiment, including VGG16-CIFAR10, VGG16-CIFAR100, ResNet20-CIFAR10, ResNet20-CIFAR100, and ResNet50-ImageNet. Among these trained models, the first four models are publicly available from the third-party GitHub repositories~\cite{cifarmodel}, and the last model is provided by PyTorch~\cite{pytorchmodel}.

Given a trained model for $N$-class classification, we perform model re-engineering to alter the trained model on two types of target problems, including binary and multi-class classification problems.
For the binary classification problem, each class of the trained model corresponds to a target problem. In total, there are $N$ target problems. A re-engineered model needs to classify whether an input belongs to the corresponding class or not.
In this scenario, VGG16-CIFAR10, VGG16-CIFAR100, ResNet20-CIFAR10, and ResNet20-CIFAR100
are altered, and there are 220 re-engineered models in total.
Due to the significant overhead of generating 1000 re-engineered models, ResNet50-ImageNet is not used here.
We count the number of removed weights and compare the accuracy of re-engineered models and trained models on target problems to validate the effectiveness of \projectName. 
Also, we compare \projectName with the state-of-the-art modularization approach~\cite{nnmodularity2022icse} to demonstrate the improvement achieved by our approach.

For the multi-class classification problem, %
a re-engineered model classifies an input into one of the concerning classes.
In this scenario, we use CIFAR-100 and ImageNet as our datasets since there are publicly available schemes for dividing them into superclasses~\cite{cifar10,imagenet_division}.
A small-size model ResNet20-CIFAR100 and a large model ResNet50-ImageNet are chosen for a more comprehensive evaluation. %
Specifically, CIFAR-100 has divided the 100 classes into 20 superclasses, each containing 5 classes with semantically similar labels~\cite{cifar10}.
For ResNet20-CIFAR100, we follow this division; thus, there are 20 target problems, each corresponding to a superclass.
For ResNet50-ImageNet, following the public division~\cite{imagenet_division}, the 1000 classes are divided into 67 superclasses, of which 3 superclasses are discarded because they contain only 1 class. 
The remaining 64 superclasses with a number of classes ranging from 2 to 119 form 64 target problems.
In total, there are 84 re-engineered models.
We count the number of removed weights and compare the accuracy of re-engineered models and trained models on target problems to validate the effectiveness of \projectName. 
Since the modularization approaches~\cite{nnmodularity2022icse,fse2020modularity} are designed for binary classification (i.e., each module performs binary classification) and cannot be applied to multi-class classification directly, we compare \projectName with the method of retraining from scratch. %

When re-engineering an original model on a target problem, we follow the settings of our baselines~\cite{ReMos,cifarmodel} to divide the target dataset into training and testing sets. The training set is used to search for a candidate, and the testing set is used to evaluate the candidate.
The major parameters in \projectName include weighting factor $\alpha$ (see Equation \ref{eq:loss}) and learning rate $\xi$ (see Equation \ref{eq:update}). 
The appropriate values of $\alpha$ and $\xi$ could vary from different trained models and are generally set to 1.0 and 0.05, respectively.
The detailed settings and their impact on model re-engineering are described in the project webpage~\cite{seam}.

\noindent \textbf{RQ2: Does reusing a re-engineered model incur less overhead than reusing the original model?} 
In this experiment, the trained models and re-engineered models from RQ1 are reused, and we compare the reuse overhead of re-engineered models with the original models.
Two metrics are used to measure the reuse overhead, including the number of floating point operations (FLOPs)~\cite{li2016pruning,luo2017thinet} and time cost for inference.
An open-source tool fvcore~\cite{fvcore} is used to calculate the FLOP.
Regarding inference time cost, an open-source tool DeepSparse~\cite{deepsparse} is used to run both original and re-engineered models and compute the inference time cost.

\noindent \textbf{RQ3: Does reusing the re-engineered model mitigate the defect inheritance?} 
In transfer learning, a pre-trained model generally has a large number of weights and classifications, and a target problem has insufficient data. Therefore, VGG16 and ResNet20 are not suitable to be transferred, and CIFAR-10 and CIFAR-100 are unsuitable for target problems.
Following the state-of-the-art approach ReMos~\cite{ReMos}, two widely-used transfer learning CNN models, ResNet18 and ResNet50, are used as trained models (i.e., the pre-trained models in transfer learning), which are trained on ImageNet and are provided by PyTorch~\cite{pytorchmodel}.
Five popular transfer learning datasets are used as target datasets, including MIT Indoor Scenes~\cite{tfdata_scenes}, Caltech-UCSD Birds~\cite{tfdata_birds}, 102 Category Flowers~\cite{tfdata_flowers}, Standford 40 Actions~\cite{tfdata_actions}, and Standford Dogs~\cite{tfdata_dogs}.

We first apply \projectName to alter the trained model on the target dataset, resulting in a re-engineered model.
Then we use the standard fine-tuning approach~\cite{transfer1, transfer2} to fine-tune the re-engineered model on the target dataset, resulting in a fine-tuned model.
We compare \projectName with two baselines, standard fine-tuning~\cite{transfer1, transfer2} and the state-of-the-art approach ReMos~\cite{ReMos}. %
Standard fine-tuning fine-tunes all of the trained model's weights on the target dataset.
ReMos first sets a trained model's weights irrelevant to the target problem to zeros, and then uses standard fine-tuning to fine-tune the sliced model on the target dataset, resulting in a fine-tuned model.
Following the setup of ReMos, we use accuracy (ACC) and defect inheritance rate (DIR) to measure and compare the effectiveness of \projectName and the baselines.
The accuracy is computed as the correct classification rate on the target dataset $D^T$:
\begin{gather}
    ACC= \frac{1}{|D^T|} \sum_{(x,y)\in D^T} \mathbbm{1}[f(x)=y].
\end{gather}
The defect inheritance rate is computed as the misclassification rate on a set of malicious inputs $S^M$:
\begin{gather}
    DIR=\frac{1}{|S^M|} \sum_{(\hat{x},y)\in S^M} \mathbbm{1}[f(\hat{x})\ne y].
\end{gather}
Same as ReMos\cite{ReMos}, open source tool \textit{advertorch}~\cite{advertorch} is used to generate $S^M$
based on the trained model and $D^T$.
We use the same parameters as ReMos when using advertorch.

In this experiment, we set the learning rate $\xi{=}0.05$ and weighting factor $\alpha{=}0.5$.
Regarding the standard fine-tuning approach and ReMos, we also use the open source project~\cite{remosproject} published by ReMos.

All the experiments are conducted on Ubuntu 20.04 server with 64 cores of 2.3GHz CPU, 128GB RAM, and NVIDIA Ampere A100 GPUs with 40 GB memory.

\subsection{Experimental Results}
\label{subsec:exp_result}
\noindent \textbf{RQ1: How effective is our model re-engineering approach in reusing trained models?}

\begin{figure}[!th]
    \centering
    \includegraphics[width=\columnwidth]{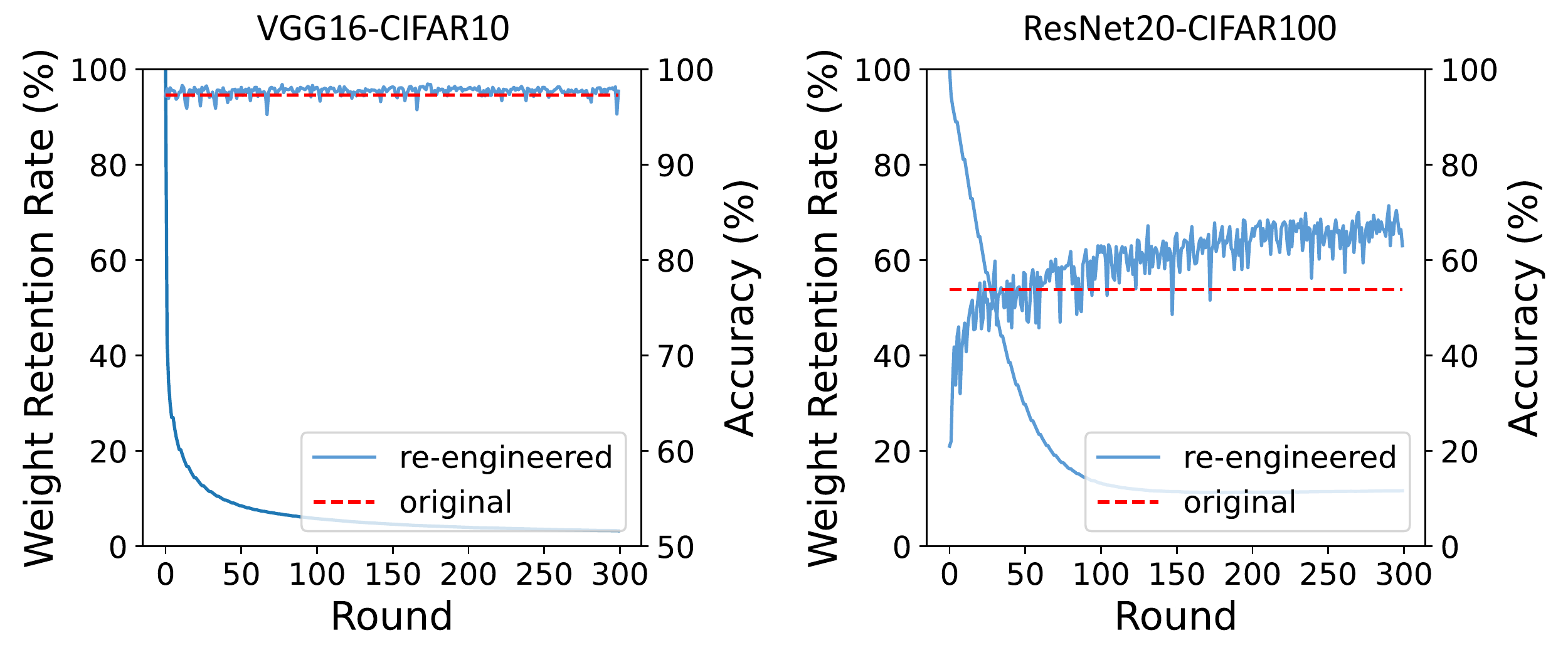}
    \caption{The convergence process of \projectName on binary (left sub-figure) and multi-class (right sub-figure) classification problems. }
    \label{fig:convergence}
\end{figure}

In this research question, we present the model re-engineering results of \projectName 
for two types of target problems (i.e., binary and multi-class classification).
Figure \ref{fig:convergence} shows the convergence process of \projectName on two types of target problems.
For instance, the left sub-figure shows the trend of weight retention rate and classification accuracy along with search rounds during re-engineering VGG16-CIFAR10 on a binary classification problem.
The weight retention rate descends quickly in the first 50 rounds and then gradually converges.
Although many weights are removed, the re-engineered model maintains a comparable accuracy to the original model.
The right sub-figure depicts the convergence process of ResNet20-CIFAR100 on a 5-class classification problem.
Similar to re-engineering VGG16-CIFAR10, the weight retention rate descends quickly in the first 100 rounds and then gradually converges.
The difference is that the accuracy of the re-engineered model may be lower than that of the original model at the beginning of search.
The reason is that the target dataset of the 5-class classification problem contains fewer samples than that of the binary classification problem (500 \textit{vs.} 10,000). Thus, the former requires more rounds to optimize the mask and head. As optimization rounds increase, the mask retains more related weights, and the head learns to classify better, so the re-engineered model can recover accuracy and eventually exceed that of the original model.
The time cost of the search varies by the models, target problems, and target datasets.
The sizes of target datasets vary from 500 to 140,000 samples.
For the binary classification problem, each round takes several seconds. 
For the multi-class classification problem, re-engineering ResNet20-CIFAR100 takes 2s per round.
For ResNet50-ImageNet, as each superclass contains a different number of classes, the time cost varies from several seconds to a few minutes per round.
In this example, re-engineering VGG16-CIFAR10 and ResNet20-CIFAR100 takes 4s and 2s per round, respectively.

Table \ref{tab:rq1_weight} shows the results regarding the number of weights for the original and re-engineered models.
For each trained model, we count the number of the original model's weights and the number of weights retained (i.e., non-zero weights) in the re-engineered model\footnote{As a head contains a negligible number of weights (e.g., 0.43\% at most) compared to the original model, the head weight count is omitted in the experiment.}.
For instance, VGG16-CIFAR10 is altered on 10 target problems, resulting in 10 re-engineered models.
The average number of weights retained (i.e., non-zero weights) in a re-engineered model is 0.62 million.
Compared to the original model having 15.25 million weights, a re-engineered model retains only 4.07\% of the original model's weights, which means that \projectName achieves a 95.93\% reduction in the number of weights.
It is worth mentioning that, for multi-class modularization, although a re-engineered model requires the classification of more classes, it still has much fewer weights than the trained model. 
For instance, a re-engineered model obtained by altering ResNet20-CIFAR100 can classify five classes; however, the re-engineered model has only an average of 0.04 million weights, and the reduction in the number of weights is 85.71\%.
The reason is that different classes may contain the same features, which means that the weights needed to identify one more class may already be included in the existing weights.
Consequently, for all six trained models, the number of weights retained in re-engineered models is significantly smaller than the number of weights in original models.
On average, for the six trained models, \projectName achieves an 89.89\% reduction in the number of weights.

\begin{table}[t]
\caption{The model re-engineering results of \projectName regarding the number of weights.}
\label{tab:rq1_weight}
\centering
\resizebox{\columnwidth}{!}{

    \begin{tabular}{ccrrr}
    \toprule
    \multirow{2}{*}{\textbf{\begin{tabular}[c]{@{}c@{}}Target\\ Problem\end{tabular}}}    & \multirow{2}{*}{\textbf{Model Name}} & \multicolumn{3}{c}{\textbf{\# Weights (million)}}                              \\ \cmidrule(lr){3-5} 
                                                                                        &                                      & \textbf{Original}      & \textbf{Re-engineered} & \textbf{Reduction(\%)} \\ \midrule \midrule
    \multirow{4}{*}{\begin{tabular}[c]{@{}c@{}}Binary\\ Classification\end{tabular}}      & VGG16-CIFAR10                        & 15.25 & 0.62                   & 95.93                  \\
                                                                                        & VGG16-CIFAR100                       &  15.29  & 1.47                   & 90.39                  \\ \cmidrule(lr){2-5} 
                                                                                        & ResNet20-CIFAR10                     & 0.27  & 0.03                   & 88.89                  \\
                                                                                        & ResNet20-CIFAR100                    & 0.28  & 0.03                   & 89.29                  \\ \midrule
    \multirow{2}{*}{\begin{tabular}[c]{@{}c@{}}Multi-class\\ Classification\end{tabular}} & ResNet20-CIFAR100                    & 0.28                   & 0.04                   & 85.71                  \\
                                                                                        & ResNet50-ImageNet                    &  25.50                      &  2.77                      & 89.16                       \\ \midrule
    \multicolumn{2}{c}{\textbf{Average}}                                                                                         & \textbf{-}             & \textbf{-}             & \textbf{89.89}              \\ \bottomrule
    \end{tabular}

}
\end{table}
\begin{table}[t]
\caption{The model re-engineering results of \projectName regarding classification accuracy.}
\label{tab:rq1_acc}
\centering
\resizebox{\columnwidth}{!}{

    \begin{tabular}{ccrrr}
    \toprule
    \multirow{2}{*}{\textbf{\begin{tabular}[c]{@{}c@{}}Target\\ Problem\end{tabular}}}    & \multirow{2}{*}{\textbf{Model Name}} & \multicolumn{3}{c}{\textbf{Avg. ACC (\%)}}                         \\ \cmidrule(lr){3-5} 
                                                                                        &                                      & \textbf{Original} & \textbf{Re-engineered} & \textbf{Increase(\%)} \\ \midrule \midrule
    \multirow{4}{*}{\begin{tabular}[c]{@{}c@{}}Binary\\ Classification\end{tabular}}      & VGG16-CIFAR10                        & 96.50             & 97.12                  & 0.62                  \\
                                                                                        & VGG16-CIFAR100                       & 86.82             & 92.93                  & 6.12                  \\ \cmidrule(lr){2-5} 
                                                                                        & ResNet20-CIFAR10                     & 95.64             & 95.81                  & 0.17                  \\
                                                                                        & ResNet20-CIFAR100                    & 83.97             & 90.92                  & 6.95                  \\ \midrule
    \multirow{2}{*}{\begin{tabular}[c]{@{}c@{}}Multi-class\\ Classification\end{tabular}} & ResNet20-CIFAR100                    & 68.29             & 82.76                  & 14.47                 \\
                                                                                        & ResNet50-ImageNet                    & 78.83             & 85.63                  & 6.80                   \\ \midrule
    \multicolumn{2}{c}{\textbf{Average}}                                                                                         & \textbf{-}        & \textbf{-}             & \textbf{5.85}         \\ \bottomrule
    \end{tabular}

}
\end{table}

Table \ref{tab:rq1_acc} shows the averaged accuracy of original and re-engineered models.
The original and re-engineered models are evaluated on the corresponding target problems. 
Again using VGG16-CIFAR10 as an example, the average accuracy of the 10 re-engineered models on the 10 target problems is 97.12\%.
The original model is also evaluated on the 10 target problems, and the average accuracy is 96.50\%. 
Compared to the original model, the re-engineered models achieve comparable accuracy on target problems, and the averaged accuracy increases by 0.62\%.
The reason for the improvement may be that model re-engineering enables the re-engineered model to fit the target problem during altering the original model.
Note that, the fitting is mainly achieved by removing irrelevant weights instead of training the weights of the original trained model.
On both binary and multi-class classification problems, for all the six trained models, re-engineered models can achieve comparable accuracy to original models, and the averaged accuracy increases by 5.85\%.
Due to space limitation, the detailed results regarding the number of weights and accuracy are available at the project webpage~\cite{seam}.

When comparing \projectName with the existing modularization approach~\cite{nnmodularity2022icse}, we directly use the open source project~\cite{modularization} published by \cite{nnmodularity2022icse}, which decomposes a trained model into modules, each for a binary classification problem.
Since tool \cite{nnmodularity2022icse} and \projectName are implemented on Keras and PyTorch, respectively, they cannot directly alter each other's trained models. 
We attempted to convert PyTorch and Keras trained models to each other; however, the conversion incurs much loss of accuracy (5\% to 10\%) due to the differences in the underlying computation of PyTorch and Keras.
To make the comparison as fair as possible, we run the modules and trained models published by \cite{nnmodularity2022icse} and compare \projectName to \cite{nnmodularity2022icse} based on the results of ResNet20-CIFAR10 and ResNet20-CIFAR100, as the two models are also used in \cite{nnmodularity2022icse}.
Specifically, we analyzed the accuracy and the number of neurons of the original models and modules (re-engineered model of \projectName). 
As modularization~\cite{nnmodularity2022icse} decomposes a CNN model mainly by removing neurons (i.e., setting neurons to zero but retaining all weights) from convolutional layers, we analyzed the number of neurons rather than the number of weights.

\begin{table}[t]
\caption{The results of modularization~\cite{nnmodularity2022icse} on binary classification problems.}
\label{tab:icse22}
\centering
\resizebox{\columnwidth}{!}{
\begin{tabular}{crrrrrr}
\toprule
\multirow{2}{*}{\textbf{Model   Name}} & \multicolumn{3}{c}{\textbf{\# Neurons (million)}}                  & \multicolumn{3}{c}{\textbf{Avg. ACC (\%)}}                  \\ \cmidrule(lr){2-4} \cmidrule(lr){5-7} 
                                       & \textbf{Original} & \textbf{Module} & \textbf{Reduction(\%)} & \textbf{Original} & \textbf{Module} & \textbf{Increase(\%)} \\ \midrule \midrule
ResNet20-CIFAR10                       &   \multirow{2}{*}{0.20}                &  0.17               & 14.33                       & 92.85             & 91.81           & -1.04                 \\
ResNet20-CIFAR100                      &                                        &  0.15               & 22.84                       & 72.92             & 59.45           & -13.47                \\ \midrule
\textbf{Average}                       & \textbf{-}         & \textbf{-}       & \textbf{18.59}              & \textbf{-}        & \textbf{-}      & \textbf{-7.25}        \\ \bottomrule
\end{tabular}
}
\end{table}

As shown in Table \ref{tab:icse22}, for both ResNet20-CIFAR10 and ResNet20-CIFAR100, a module retains fewer neurons than the original model; however, the number of neurons in a module is reduced by only 18.59\% on average. In addition, a module retains all the weights of the convolutional layers.
Regarding accuracy, modules achieve a lower accuracy than the trained models on target problems, and the accuracy of a module reduces by 7.25\% on average. 
Compared to modularization~\cite{nnmodularity2022icse}, model re-engineering can remove a large number of weights without impairing the accuracy.
A major reason for the improvement of \projectName over \cite{nnmodularity2022icse} is that \projectName identifies the target problem-related weights more accurately.
\projectName is a search-based approach that identifies the target problem-related weights directly based on the classification accuracy, while \cite{nnmodularity2022icse} identifies the target problem-related weights and neurons based on the neuron activation that indirectly correlates with the accuracy.

\begin{table}[t]
\caption{The model re-engineering and model retraining results on multi-class classification.}
\label{tab:rq1_retraining}
\centering
\resizebox{0.8\columnwidth}{!}{

    \begin{tabular}{crrr}
    \toprule
    \multirow{2}{*}{\textbf{Model Name}} & \multicolumn{3}{c}{\textbf{Avg. ACC (\%)}}                           \\ \cmidrule(lr){2-4} 
                                        & \textbf{Retrained} & \textbf{Re-engineered} & \textbf{Increase (\%)} \\ \midrule \midrule
    ResNet20-CIFAR100                    & 77.15              & 82.76                  & 5.61                   \\
    ResNet50-ImageNet                    & 75.56              & 85.63                  & 10.08                  \\ \midrule
    \textbf{Average}                     & \textbf{-}         & \textbf{-}             & \textbf{7.84}          \\ \bottomrule
    \end{tabular}
}
\end{table}
We also compare \projectName to model retraining on multi-class classification problems.
Model retraining reuses the architecture and hyperparameters of the trained model to retrain a new model from scratch on the target dataset.
As both model re-engineering and retraining alter/train the same model (architecture) on the same target problem, while the latter may fit more slowly and even run several times, the time cost of the retraining would be higher than that of re-engineering.
Regarding accuracy, as shown in Table \ref{tab:rq1_retraining},
re-engineered models outperform retrained models for both ResNet20-CIFAR100 and ResNet50-ImageNet, and the average improvement is 7.84\%.
The reason for the improvement of model re-engineering may be the difference in the amount of data.
The original model is trained on a large-scale dataset, while the retrained model is trained on a small-scale target dataset.
The model re-engineering alters the original model to fit the target problem; thus, the re-engineered model achieves higher accuracy than the retrained model.

\begin{tcolorbox}[left=2pt,right=2pt,top=2pt,bottom=2pt]
On average, a re-engineered model contains 89.89\% fewer weights than the original model but outperforms the original model in accuracy by 5.85\%.
\end{tcolorbox}

\begin{table}[t]
\caption{The number of FLOPs required by original and re-engineered models.}
\label{tab:rq2_flop}
\centering
\resizebox{\columnwidth}{!}{

    \begin{tabular}{ccrrr}
    \toprule
    \multirow{2}{*}{\textbf{\begin{tabular}[c]{@{}c@{}}Target\\ Problem\end{tabular}}}    & \multirow{2}{*}{\textbf{Model Name}} & \multicolumn{3}{c}{\textbf{FLOPs (million)}}                                     \\ \cmidrule(lr){3-5} 
                                                                                        &                                      & \textbf{Original}       & \textbf{Re-engineered} & \textbf{Reduction (\%)} \\ \midrule
    \multirow{4}{*}{\begin{tabular}[c]{@{}c@{}}Binary\\ Classification\end{tabular}}      & VGG16-CIFAR10                        & 314.28 & 75.53                  & 75.97                   \\
                                                                                        & VGG16-CIFAR100                       &   314.33                      & 111.35                 & 64.58                   \\ \cmidrule(lr){2-5} 
                                                                                        & ResNet20-CIFAR10                     & 41.22  & 9.34                   & 77.35                   \\
                                                                                        & ResNet20-CIFAR100                    & 41.22                        & 9.36                   & 77.30                   \\ \midrule
    \multirow{2}{*}{\begin{tabular}[c]{@{}c@{}}Multi-class\\ Classification\end{tabular}} & ResNet20-CIFAR100                    & 41.22                   & 9.76                   & 76.32                   \\
                                                                                        & ResNet50-ImageNet                    &  4111.53                       &  955.85                      &  76.75                       \\ \midrule
    \multicolumn{2}{c}{\textbf{Average}}                                                                                         & \textbf{-}              & \textbf{-}             & \textbf{74.71}               \\ \bottomrule
    \end{tabular}

}
\end{table}
\begin{table}[t]
\caption{The inference time cost required by original and re-engineered models.}
\label{tab:rq2_time}
\centering
\resizebox{\columnwidth}{!}{

    \begin{tabular}{ccrrr}
    \toprule
    \multirow{2}{*}{\textbf{\begin{tabular}[c]{@{}c@{}}Target\\ Problem\end{tabular}}}    & \multirow{2}{*}{\textbf{Model Name}} & \multicolumn{3}{c}{\textbf{Time Cost (ms/batch)}}                    \\ \cmidrule(lr){3-5} 
                                                                                        &                                      & \textbf{Original} & \textbf{Re-engineered} & \textbf{Reduction (\%)} \\ \midrule
    \multirow{4}{*}{\begin{tabular}[c]{@{}c@{}}Binary\\ Classification\end{tabular}}      & VGG16-CIFAR10                        & 6.82              & 3.79                   & 44.43                   \\
                                                                                        & VGG16-CIFAR100                       & 6.37              & 5.55                   & 12.87                   \\ \cmidrule(lr){2-5} 
                                                                                        & ResNet20-CIFAR10                     & 1.40              & 0.74                   & 47.14                   \\
                                                                                        & ResNet20-CIFAR100                    & 1.43              & 0.64                   & 55.24                   \\ \midrule
    \multirow{2}{*}{\begin{tabular}[c]{@{}c@{}}Multi-class\\ Classification\end{tabular}} & ResNet20-CIFAR100                    & 1.45              & 0.61                   & 57.93                   \\
                                                                                        & ResNet50-ImageNet                    &  64.19                 & 40.55                       &  36.83                       \\ \midrule
    \multicolumn{2}{c}{\textbf{Average}}                                                                                         & \textbf{-}        & \textbf{-}             & \textbf{42.41}               \\ \bottomrule
    \end{tabular}
}
\end{table}

\noindent \textbf{RQ2: Does reusing a re-engineered model incur less overhead than reusing the original model?}
\label{subsec:results:rq2}

One of the benefits of model re-engineering is to reduce the reuse overhead.
As mentioned in Section \ref{subsec:setup}, the number of FLOPs and inference time cost are used to measure the reuse overhead.
We evaluated the original and re-engineered models from RQ1 on the two metrics to answer this research question.

Table \ref{tab:rq2_flop} shows the number of FLOPs required by the original and re-engineered models to classify an image with resolution $32{\times}32$, respectively.
Note that, following the related work~\cite{pruning_hansong,louizos2018learning}, when computing the number of FLOPs required by a re-engineered model with a sparse weight matrix, only the computations involved in non-zero weights are considered.
For instance, despite having the same number of weights as the original model, a re-engineered model obtained by altering VGG16-CIFAR10 has 95.93\% (see Table \ref{tab:rq1_weight}) of its weights set to zero.
As the calculations associated with these zero weights can be eliminated by special libraries~\cite{han2016eie}, the calculations associated with these weights are not considered when calculating FLOPs.
VGG16-CIFAR10 requires 314.28 million FLOPs, while the average number of FLOPs required by a re-engineered model is 75.53 million. 
\projectName achieves 75.97\% reduction in terms of FLOPs.
On average, for the six trained models, \projectName reduces the FLOPs by 74.71\%.

To verify that the reduction in the number of FLOPs can reduce the inference time cost, the open-source library DeepSparse~\cite{deepsparse} is used to deploy and run the original and re-engineered models. 
Given an input with batch size 16, each re-engineered model or original model classifies the input 200 times, and the average time cost of classification is used to measure the inference time cost of a re-engineered model or an original model.
Table \ref{tab:rq2_time} shows the average inference time cost of each trained model and its corresponding re-engineered models. 
For instance, the inference time cost of VGG16-CIFAR10 is 6.82ms/batch, which means that VGG16-CIFAR10 requires 6.82ms to classify an input with batch size 16. The re-engineered model obtained by altering VGG16-CIFAR10 incurs an average of 3.79ms/batch inference time cost. The reduction in inference time cost is 44.43\% (calculated by $(1-3.79/6.82)*100$).
For all the six trained models, \projectName achieves an average of 42.41\% reduction in inference time cost, which demonstrates that the reduction in the number of weights and FLOPs can reduce the inference time cost.

FLOP focus on the computation of neural network layers containing weights.
While apart from the layers containing weights, the time cost for inference also involves other operations, such as activation functions, dropout, tensor reshape, and so on.  
Therefore, the reductions in the number of FLOPs and the time cost differ. %

\begin{tcolorbox}[left=2pt,right=2pt,top=2pt,bottom=2pt]
Reusing a re-engineered model incurs less reuse overhead than reusing an original model, while achieving even higher accuracy in inference than the original model.
\end{tcolorbox}

\noindent \textbf{RQ3: Does reusing the re-engineered model mitigate the defect inheritance?}

\begin{figure}
    \centering
    \includegraphics[width=\columnwidth]{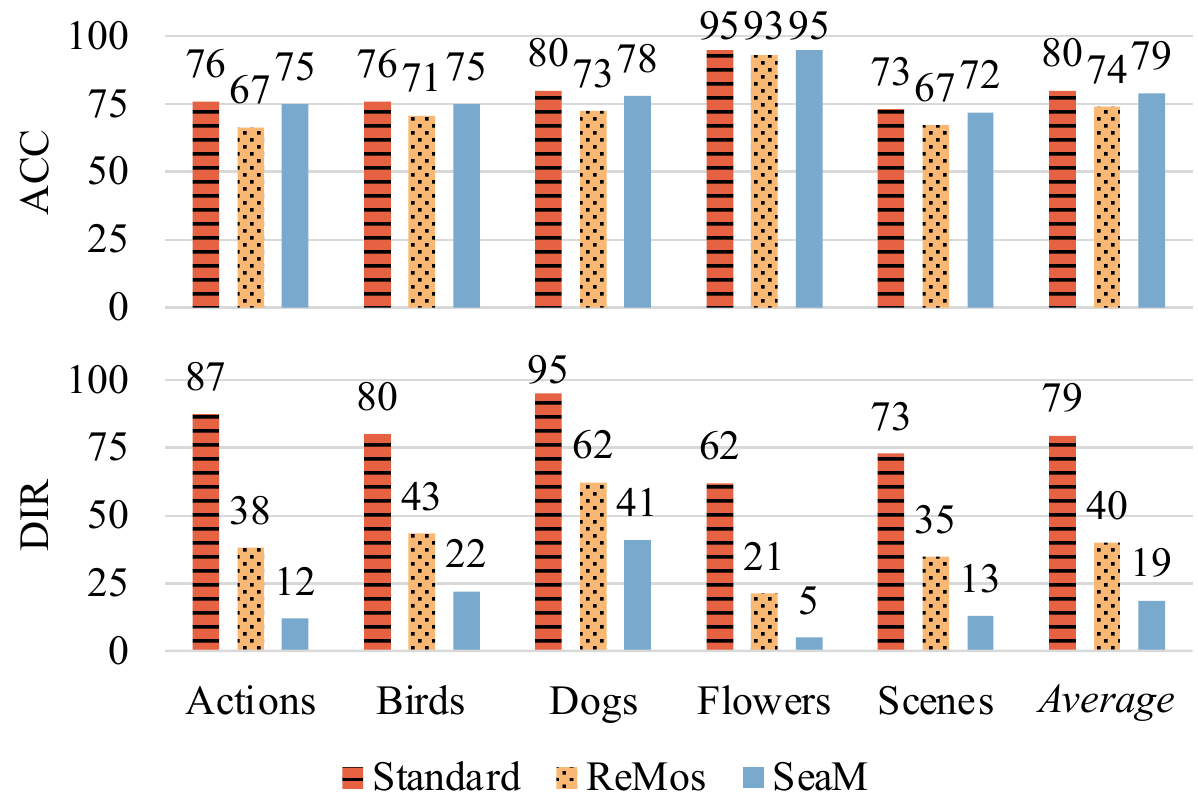}
    \caption{The accuracy (ACC) and defect inheritance rate (DIR) on ResNet18.}%
    \label{fig:adv_resnet18}
\end{figure}

\begin{figure}
    \centering
    \includegraphics[width=\columnwidth]{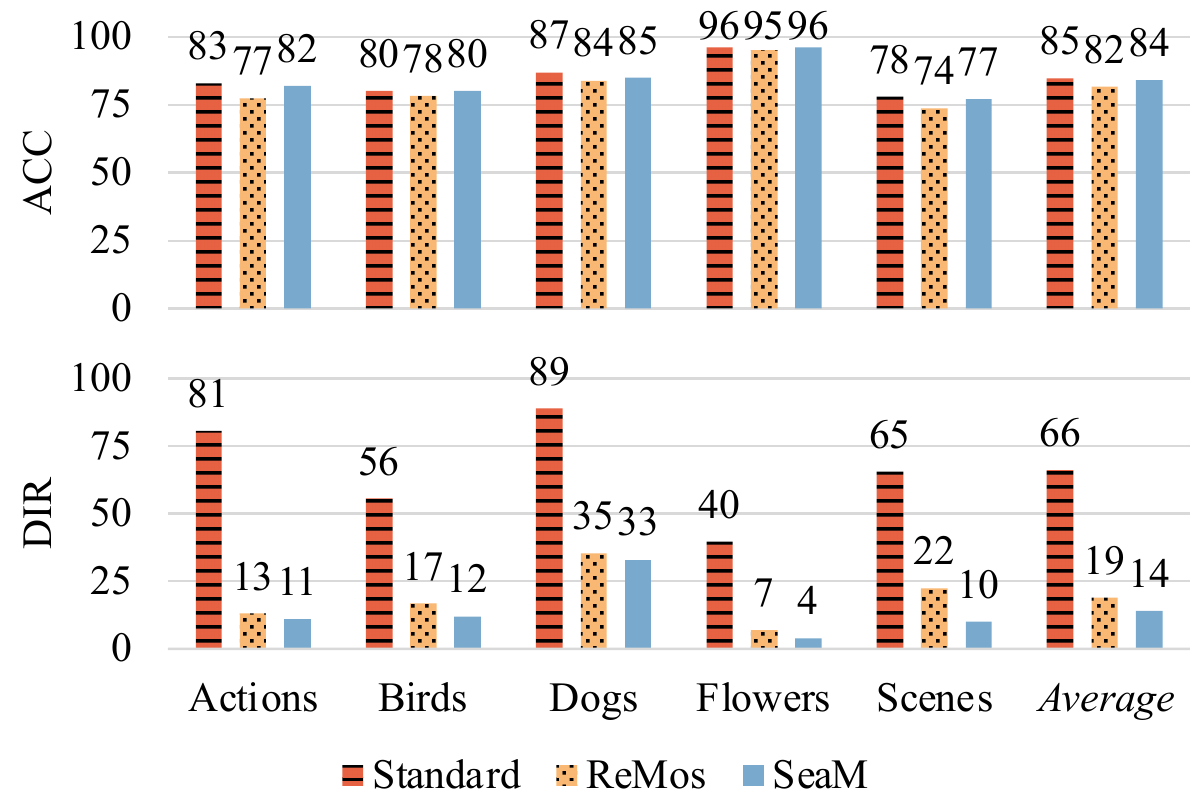}
    \caption{The accuracy (ACC) and defect inheritance rate (DIR) on ResNet50.}
    \label{fig:adv_resnet50}
\end{figure}

In this experiment, following ReMos, we evaluate \projectName and the baselines using two metrics, i.e., accuracy (\textit{ACC}) and defect inheritance rate (\textit{DIR}).
Figure \ref{fig:adv_resnet18} and Figure \ref{fig:adv_resnet50} show the results of ResNet18 and ResNet50 on all of five datasets. 
In each figure, the first row displays the accuracy (\textit{ACC}) of the fine-tuned model, and the second row displays the defect inheritance rate (\textit{DIR}) of the fine-tuned model.
On average, for ResNet18, the ACC and DIR achieved by the standard fine-tuning, ReMos, and \projectName are (80\%, 79\%), (74\%, 40\%), and (79\%, 19\%), respectively. For ResNet50, the ACC and DIR are (85\%, 66\%), (82\%, 19\%), (84\%, 14\%), respectively.

The standard fine-tuning approach achieves higher accuracy on the target datasets than \projectName and ReMos; however, the cost is much higher DIRs.
Compared to the standard fine-tuning approach, both \projectName and ReMos can achieve lower DIRs at the cost of a small accuracy loss, indicating that removing the weights that are not relevant to the target dataset can reduce DIRs and improve the robustness of the fine-tuned model.
Overall, for the two models on five datasets, the averaged DIRs for fine-tuning the re-engineered model and fine-tuning the original model (i.e., standard fine-tuning approach) are 16\% and 73\%, respectively.
The reduction in DIR is 57\%, demonstrating the effectiveness of \projectName in reducing defect inheritance.

Compared to ReMos, \projectName can achieve lower DIRs and higher ACC.
For instance, for ResNet18, the average DIRs achieved by \projectName and ReMos are 19\% and 40\%, respectively. The DIR achieved by \projectName is roughly half of that achieved by ReMos.
Regarding ACC, the average ACC achieved by \projectName and ReMos is 79\% and 74\%, respectively. 
Overall, for the two models on five datasets, the average DIRs and ACC for \projectName and ReMos are (16\%, 82\%) and (29\%, 78\%), respectively.
\projectName is 13\% lower and 4\% higher than ReMos in terms of DIR and ACC, respectively.
The reason for the improvement in DIR achieved by \projectName is the considerable reduction in the number of weights.
ReMos removes only 10\% and 3\% weights for ResNet18 and ResNet50, respectively.
Compared to ReMos, \projectName can remove more irrelevant weights. 
The reduction in the number of weights is about 50\% for both ResNet18 and ResNet50.

\begin{figure}
    \centering
    \includegraphics[width=\columnwidth]{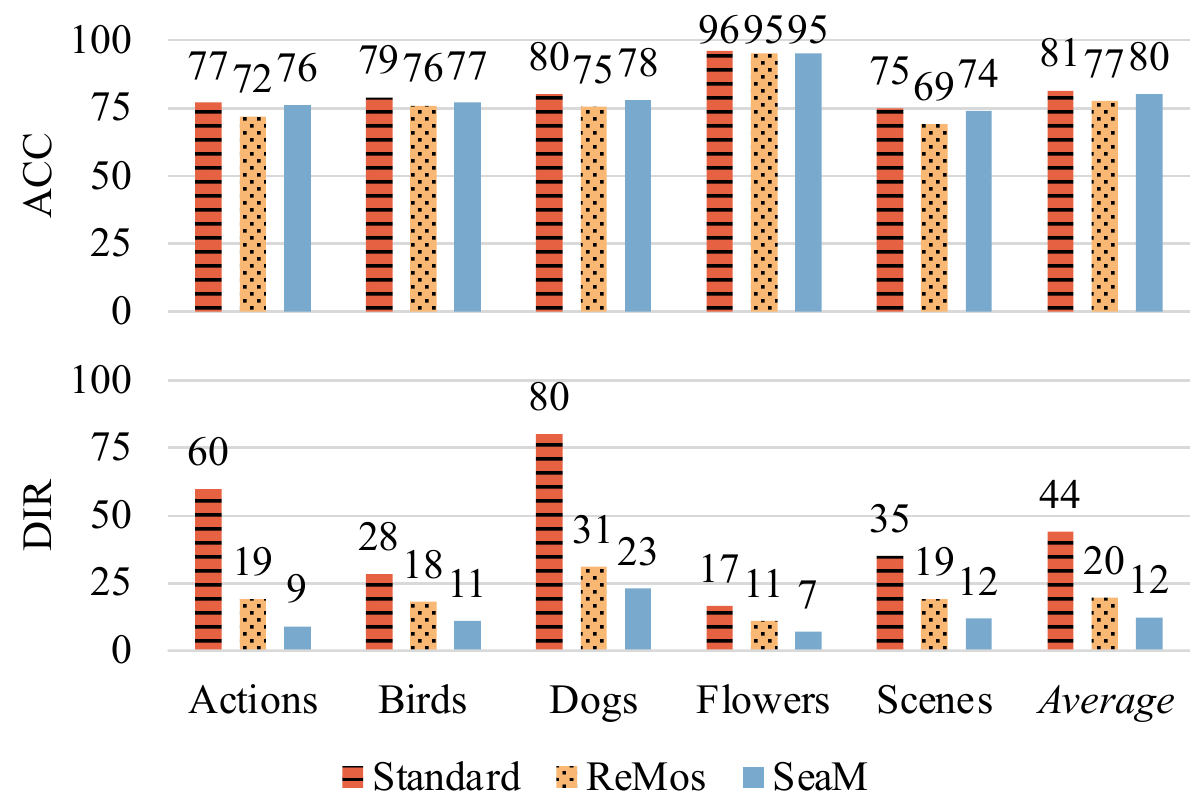}
    \caption{The accuracy (ACC) and defect inheritance rate (DIR) on ResNet18 with Dropout layers.}
    \label{fig:adv_resnet18_dropout}
\end{figure}

It is worth mentioning that there are some differences between the results shown in Figure \ref{fig:adv_resnet18} and Figure \ref{fig:adv_resnet50} and the results shown in ReMos~\cite{ReMos}, especially in terms of DIR.
For instance, for ResNet18, the average DIRs achieved by ReMos shown in \cite{ReMos} and our work are 15\% and 40\%, respectively. 
The reason for the differences is that ReMos uses additional Dropout layers for fine-tuning while ours does not.
To make a more comprehensive comparison of ReMos and \projectName, we follow the experimental setup of ReMos~\cite{ReMos} and plot the results on ResNet18 in Figure \ref{fig:adv_resnet18_dropout}.
As shown in Figure \ref{fig:adv_resnet18_dropout}, after adding Dropout layers, both \projectName and the baselines achieve better results, as Dropout layers help increase the robustness of models. 
The average DIRs achieved by the standard fine-tuning approach, ReMos, and \projectName are 44\%, 20\%, and 12\%, respectively. 
Consistent with the above conclusion, our approach can outperform %
ReMos.
Moreover, we observe that the DIRs of ResNet50 are lower than that of ResNet18. 
The reason for this could be that the increased number of weights helps increase the robustness.
This observation aligns with the prior works~\cite{ReMos,adv_attack}.

\begin{tcolorbox}[left=2pt,right=2pt,top=2pt,bottom=2pt]
Overall, \projectName inherits much fewer defects compared to standard fine-tuning and the state-of-the-art approach. %
\end{tcolorbox}

\section{Threats to Validity}
\label{sec:threats}
\textbf{External validity:}
Threats to external validity relate to the generalizability of our results. 
While the notion of re-engineering a trained model to improve its reusability %
is general, we have only evaluated our approach on CNN models in this paper.
The effectiveness on other types of DNNs, such as LSTM and transformer, remains to be evaluated. 
However, during the search, the objects removed are weights, not CNN-specific structures such as convolutional kernels.
Also, the search is guided by the classification accuracy and the number of retained weights.
Therefore, the principles of our proposed approach are not specific to CNN and %
are applicable to other types of DNNs as well. We will further investigate it in our future work.

\textbf{Internal validity:}
An internal threat comes from the choice of trained models and datasets. To mitigate this threat, we use four representative trained CNN models and evaluate \projectName on eight well-organized and widely-used datasets.

\textbf{Construct validity:}
A threat relates to the suitability of our evaluation metrics. 
Evaluating the quality of DNN models remains an open problem.
Measuring only the misclassification rate of the adversarial samples may not be comprehensive enough. 
However, the misclassification rate of adversarial samples is a representative metric and has also been widely used in  related work~\cite{ReMos, defect2}.

\section{Related Work}
\label{sec:related}
\textit{Reusing trained DNN models:} 
Our work is related to reusing DNN models, including direct reuse~\cite{icse21discriminiate, ji2018model} and transfer learning~\cite{transfer_nips,guo2019spottune}.
The work related to direct reuse recommends a trained model for developers and allows developers to reuse the model on the target problem directly.
For instance, SDS~\cite{icse21discriminiate} evaluates trained models using a few efficient test data that could discriminate multiple trained models and then recommends the best one to reuse.
Transfer learning techniques reuse a model trained to solve a similar problem and fine-tune the reused model on the target problem.
For instance, ResNet~\cite{resnet} trained on ImageNet for 1000-class classification is widely reused to develop new models for various target problems by fine-tuning its weights on the target datasets~\cite{kornblith2019better,guo2019spottune}.
The techniques mentioned above support model reuse; however, they reuse the entire trained model or the vast majority of model's weights.
In contrast, this work allows developers to reuse only the target problem-related weights, thus reducing reuse overhead and defect inheritance.

\textit{DNN modularization and slicing:}
Similar to our work, DNN modularization~\cite{nnmodularity2022icse, fse2020modularity} and slicing~\cite{ReMos} attempt to reuse part of trained models. 
For instance, DNN modularization~\cite{nnmodularity2022icse, fse2020modularity} decomposes a trained model into modules based on neuron activation~\cite{neuron_activation,li2019structural}. A module retains part of trained model's neurons and can be reused to solve a binary classification problem.
Relying on neuron coverage~\cite{neuron_activation,li2019structural}, DNN slicing~\cite{ReMos} removes irrelevant weights and reuses the slice with relevant weights for fine-tuning. 
Compared to DNN modularization and slicing, our work is search-based model re-engineering, which can remove much more irrelevant weights and hence reduce more reuse overhead and defect inheritance. Our previous work CNNSplitter~\cite{qi2022patching} concerns the modularization of CNN models through searching with genetic algorithms and fixing the weakness of a model by replacing the corresponding part with a better module. In contrast, this work can realize the modularization of general neural network models and the searching algorithm is more efficient.

\textit{DNN pruning:}
Iterative magnitude pruning~\cite{pruning_hansong, lottery, rosenfeld2021predictability} is one of the mainstream network pruning techniques, which prunes part of weights that are not important for the original problem to reduce the computational overhead required by inference on the original problem.
Our work removes part of weights that are irrelevant to a target problem to reduce reuse overhead and defect inheritance on the target problem.
Apart from their differences in objectives, iterative magnitude pruning compresses a model by repeatedly removing unimportant weights and retraining the retained weights over several rounds, while \projectName removes irrelevant weights without changing retained weights.


\section{Conclusion}
\label{sec:conclusion}
In this work, we propose the notion of \textit{model re-engineering}, which re-engineers a trained DNN model to improve its reusability.
Based on the notion, we propose a search-based model re-engineering approach named \projectName, which can re-engineer a trained model by removing many irrelevant weights.
Extensive experiments with four representative CNN models on eight widely-used datasets demonstrate the effectiveness of \projectName in reusing trained models as well as reducing reuse overhead and defect inheritance.

Our source code and experimental data are available at: \textbf{\url{https://github.com/qibinhang/SeaM}}.

\section*{Acknowledgement}
This work was supported partly by National Natural Science Foundation of China under Grant Nos.(61932007, 61972013, 62141209, 62202026) and Australian Research Council (ARC) Discovery Project DP200102940 and sponsored by Huawei Innovation Research Plan.

\balance

\bibliographystyle{IEEEtran}
\bibliography{reference}

\end{document}